# Dynamic correlations with time dependent quantum Monte Carlo

Ivan P. Christov

Physics Department, Sofia University, 1164 Sofia, Bulgaria

**Abstract**

In this paper, we solve quantum many-body problem by propagating ensembles of trajectories and guiding waves in physical space. We introduce the "effective potential" correction within the recently proposed time-dependent quantum Monte Carlo methodology to incorporate the nonlocal quantum correlation effects between the electrons. The associated correlation length is calculated by adaptive kernel density estimation over the walker distribution. The general formalism is developed and tested on one-dimensional Helium atom in laser field of different intensity and carrier frequency. Good agreement with exact results for the atomic ionization is obtained.

PACS: 31.25.-v; 02.70.Ss

# 1. Introduction

Understanding the microscopic mechanisms responsible for the motion of electrons on the atomic scale is of primary importance because it would allow one to control processes like excitation of atoms and molecules into a desired quantum state, manipulation of chemical bonds, electron transport in nano-electronic circuits, etc. The increasing interest in this direction is stimulated by the new experimental opportunities owing to the recent advent of laser pulses of a few tens to hundreds attosecond duration which opens the way to probing events on the time scale of electronic motion in atoms and solid state [1,2]. Measurements with attosecond time resolution that involve excitation and relaxation of many-electron systems have been reported [3,4]. The adequate description of such attosecond experiments requires non-perturbative time-dependent analysis of the collective motion in many-electron quantum systems. In this area there is promise for theory coupled with high-end computation to achieve efficiency sufficient to treat ultrafast processes in clusters and nanostructures. The standard techniques for ground-state calculations using Gaussian- or Slater-type orbitals are not appropriate for describing the interaction of atoms and molecules with strong ultrafast laser fields where significant deformation of the electron cloud occurs. On the other hand, there has been noticeable development of time-dependent density functional theory (TDDFT) where the generally unknown exchange-correlation potential has to be approximated for each case [5-7] since there is no systematic way to improve that potential. Several failures of TDDFT in ultrafast regime have been observed [8-12]. Other recent methods to treat the many-electron quantum dynamics include the multi-configuration time dependent Hartree-Fock [13-15] method and the time dependent configuration interaction [16,17] method. Both methods however use large number of Slater determinants to be propagated in time that involves calculation of numerous integrals over the configurations, which requires large computer resources [18]. One of the reasons why the multi-configuration methods are computationally-demanding is that they use multiple waves even for highly accelerated electrons whose de Broglie wavelength is rather small and they can be treated as classical particles.

Recently an alternative approach, the time-dependent quantum Monte Carlo (TDQMC) was proposed which employs both particles and quantum waves in order to describe the ground state and the time evolution of many-electron systems [19,20]. Time-dependent quantum Monte Carlo is reminiscent of the conventional diffusion quantum Monte Carlo (DMC) [21,22] in that both methods use walkers and guiding waves. In TDQMC the electron is described statistically as an ensemble of walkers in physical space where each walker is guided by a separate time-dependent de Broglie-Bohm pilot wave. One major difference between DMC and TDQMC is that in the latter the guiding waves obey a set of coupled time-dependent 3D Schrödinger equations. This allows the consistent time evolution of these waves together with the motion of the walkers to be calculated. Unlike in other methods that use quantum trajectories to describe complex systems (see e.g. [23]), the calculation of the Bohmian quantum potential is avoided in TDQMC. In order to account for the electron-electron interaction TDQMC uses explicit Coulomb potentials instead of exchange-correlation potentials. Previous calculations have shown that the ground state obtained in TDQMC is very close to the full configuration interaction result [20]. In this paper we further explore the TDQMC method by introducing the "effective potential" concept and the relevant correlation length, which better describe the nonlocal correlation between the electrons and reconciles the ultra-correlated and non-correlated limits within the TDQMC framework. We next apply the method to one-dimensional Helium atom in an external field of different intensity and frequency.

## 2. Ultra-correlated versus uncorrelated quantum dynamics

Within the fixed-nuclei approximation, the *N*-electron system is described by the many-body Schrödinger equation:

$$i\hbar \frac{\partial}{\partial t}\Psi(\mathbf{R},t) = -\frac{\hbar^2}{2m}\nabla^2\Psi(\mathbf{R},t) + V(\mathbf{R})\Psi(\mathbf{R},t) \;, \tag{1}$$

where $\mathbf{R} = (\mathbf{r}_1,...,\mathbf{r}_N)$ is a 3*N* dimensional vector in configuration space which specifies the coordinates of *N* electrons, and $\nabla = (\nabla_1, \nabla_2,...,\nabla_N)$. The Hamiltonian in Eq. (1) is inseparable in the electron coordinates, due to the electron-electron interaction:

$$V(\mathbf{r}_1,...,\mathbf{r}_N) = V_{e-n}(\mathbf{r}_1,...,\mathbf{r}_N) + V_{e-e}(\mathbf{r}_1,...,\mathbf{r}_N) + V_{ext}(\mathbf{r}_1,...,\mathbf{r}_N,t) \tag{2}$$

$$= \sum_{k=1}^{N} V_{e-n}(\mathbf{r}_k) + \sum_{k>l}^{N} V_{e-e}(\mathbf{r}_k - \mathbf{r}_l) + V_{ext}(\mathbf{r}_1,...,\mathbf{r}_N,t) \;,$$

where the many-body classical potential in Eq. (2) is a sum of electron-nuclear, electron-electron, and external potentials. In the TDQMC methodology the many-body quantum system is described by ensembles of point particles (walkers) and the corresponding ensembles of guiding waves. Note that in TDQMC the guiding waves preserve their standard statistical interpretation of quantum mechanics. Formally, the particle concept is introduced in TDQMC by representing the wave-function as a polar decomposition:

$$\Psi(\mathbf{R},t) = R(\mathbf{R},t)\exp[iS(\mathbf{R},t)/\hbar], \qquad (3)$$

where $R(\mathbf{R},t)$ and $S(\mathbf{R},t)$ are real-valued functions of space and time. Then, inserting Eq. (3) into Eq. (1) and separating the real and imaginary parts, we obtain the generalized Hamilton-Jacobi equation [24]:

$$\frac{\partial S(\mathbf{R},t)}{\partial t} + \frac{1}{2m}\sum_{i=1}^{N}[\nabla_i S(\mathbf{R},t)]^2 + Q(\mathbf{R},t) + V(\mathbf{R},t) = 0 \qquad (4)$$

and the corresponding continuity equation for the probability density in configuration space. In Eq. (4) $Q(\mathbf{R},t)$ is the many-body quantum potential. In the stochastic interpretation of quantum mechanics [25], statistical ensembles of particles are used where the quantum system tends to equilibrium via fluctuations, where at equilibrium the particle distribution function $P(\mathbf{R},t)$ obeys $P(\mathbf{R},t) = |\Psi(\mathbf{R},t)|^2$ [26,27]. In order to transform the stochastic theory from configuration space to physical space we reduce the 3$N$-dimensional Schrödinger equation to a set of coupled 3-dimensional Schrödinger equations for the separate guiding waves using a factorization of the amplitude of the many-body wave-function in Eq. (3) [20]:

$$R(\mathbf{r}_1,...,\mathbf{r}_N,t) = R_1(\mathbf{r}_1,t)...R_N(\mathbf{r}_N,t) \tag{5}$$

This factorization is equivalent to a separable *N*-body density of particles. In this case the many-body quantum potential $Q(\mathbf{R},t)$ reduces to a sum of the quantum potentials for the separate particles. In this way the factorization in Eq. (5) allows us to ascribe a separate 3D time-dependent Schrödinger equation to each individual wavefunction $\varphi_i^k(\mathbf{r}_i,t)$ which guides the *k*-th walker from the *i*-th electron ensemble [20]:

$$i\hbar\frac{\partial}{\partial t}\varphi_i^k(\mathbf{r}_i,t) = \left[-\frac{\hbar^2}{2m}\nabla_i^2 + V_{e-n}(\mathbf{r}_i) + \sum_{j\neq i}^{N} V_{e-e}[\mathbf{r}_i - \mathbf{r}_j^k(t)] + V_{ext}(\mathbf{r}_i,t)\right]\varphi_i^k(\mathbf{r}_i,t), \tag{6}$$

where $\mathbf{r}_j^k(t)$ is the trajectory of the *k*-th walker. It can be seen that the pairwise Coulomb interaction between the Bohmian particles is accounted for quantum-mechanically in Eq. (6). However, some important quantum correlation effects which the many-body wavefunction introduces on the particle motion are still missing in Eq. (6). For example, the exchange interaction between parallel spin electrons is accounted for in TDQMC by representing the many-body quantum state as an antisymmetrized product (Slater determinant):

$$\Psi(\mathbf{r}_1,\mathbf{r}_2,...,\mathbf{r}_N,t) = A\prod_{i=1}^{N}\varphi_i(\mathbf{r}_i,t), \tag{7}$$

which is next substituted into the guidance equation for the velocity of the Bohmian walkers:

$$\mathbf{v}(\mathbf{r}_i^k) = \frac{\hbar}{m} \text{Im} \left[ \frac{1}{\Psi(\mathbf{r}_1,...,\mathbf{r}_N,t)} \nabla_i \Psi(\mathbf{r}_1,...,\mathbf{r}_N,t) \right]_{\mathbf{r}_j = \mathbf{r}_j^k(t)} \quad (8)$$

In this way the time dependent nodes of the many-body quantum state rule the motion of the walkers. Additionally the walkers experience a random drift that thermalizes their distribution, where the walker's coordinate is updated in time according to:

$$d\mathbf{r}_i^k = \mathbf{v}(\mathbf{r}_i^k) dt + \mathbf{\eta} \sqrt{\frac{\hbar}{m} dt} \quad , \quad (9)$$

where $\mathbf{\eta}$ is a vector random variable with zero mean whose variance decreases as we approach the ground state of the system.

In fact, the set of Schroedinger equations (6) describes the evolution of coupled guiding waves in physical space which are correlated due to the Coulomb interaction between the electrons, while the original Schroedinger equation (Eq. (1)) describes the evolution of many-body quantum wave in configuration space that may involve additional nonlocal quantum correlations between the electrons. With other words, if we ignore for a moment the exchange interaction, from Eq. (6) and Eq. (8) it follows that each point in configuration space which represents the instantaneous coordinates of all electrons, moves independently from the other points of the ensemble in that space. This is a direct consequence of the factorization done in Eq. (5) which reduced the 3N-dimensional quantum diffusion to 3 dimensions. This case can be labeled "ultra-correlated" because it essentially overestimates the pairwise Coulomb interaction between the walkers. In order to further clarify this point let us consider the opposite, uncorrelated case, where all electrons move in the mean-field Hartree potential. The time dependent Hartree approximation is obtained by performing a full factorization of the many-body wavefunction $\Psi(\mathbf{r}_1,...,\mathbf{r}_N,t) = \varphi_1(\mathbf{r}_1,t)...\varphi_N(\mathbf{r}_N,t)$, which from Eq. (1) yields:

$$i\hbar\frac{\partial}{\partial t}\varphi_i(\mathbf{r}_i,t) = \left[-\frac{\hbar^2}{2m}\nabla_i^2 + V_{e-n}(\mathbf{r}_i) + \sum_{j\neq i}^{N}\int d\mathbf{r}_j V_{e-e}(\mathbf{r}_i-\mathbf{r}_j)|\varphi_j(\mathbf{r}_j,t)|^2 + V_{ext}(\mathbf{r}_i,t)\right]\varphi_i(\mathbf{r}_i,t) \qquad (10)$$

If we now substitute the probability density for the *i*-th electron with its representation as a sum of delta-functions over the ensemble of *M* Monte-Carlo sample points (walkers) with definite trajectories:

$$|\varphi_j(\mathbf{r}_j,t)|^2 = \frac{1}{M}\sum_{k=1}^{M}\delta\left[\mathbf{r}_j-\mathbf{r}_j^k(t)\right] \quad , \qquad (11)$$

we obtain form Eq. (10):

$$i\hbar\frac{\partial}{\partial t}\varphi_i(\mathbf{r}_i,t) = \left[-\frac{\hbar^2}{2m}\nabla_i^2 + V_{e-n}(\mathbf{r}_i) + \sum_{j\neq i}^{N}\overline{V}_{e-e}[\mathbf{r}_i-\mathbf{r}_j(t)] + V_{ext}(\mathbf{r}_i,t)\right]\varphi_i(\mathbf{r}_i,t), \qquad (12)$$

where:

$$\bar{V}_{e-e}[\mathbf{r}_i - \mathbf{r}_j(t)] = \frac{1}{M}\sum_{k=1}^{M} V_{e-e}[\mathbf{r}_i - \mathbf{r}_j^k(t)] \tag{13}$$

is the average electron-electron potential seen by the *i*-th electron due to all Bohmian walkers which approximate the *j*-th electron. The walker motion in not correlated in this approximation because all guiding waves in Eq. (12) depend on almost the same average electron-electron potential of Eq. (13). Since the ultra-correlated and the uncorrelated cases (Eq. (6) and Eq. (12), respectively) are not related in an obvious way, it is useful to introduce an approach where the electron correlation can be "switched on" so that the pairwise Coulomb interaction between the walkers incrementally replaces the mean-field Hartree potential. In this way the role of the quantum fluctuations that are responsible for the long-range electron correlation in configuration space can be estimated. Here we use a simple heuristic approach where we replace the delta-function approximation for the density in Eq. (11) by a smoothed interpolation (kernel density estimation) with Gaussians, centered at the positions of the Bohmian walkers:

$$|\varphi_j(\mathbf{r}_j,t)|^2 = \frac{1}{Z_j}\sum_{k=1}^{M}\exp\left(-\frac{|\mathbf{r}_j - \mathbf{r}_j^k(t)|^2}{\sigma_j^k\left(\mathbf{r}_j^k,t\right)^2}\right), \tag{14}$$

where the width of the Gaussians $\sigma_j^k\left(\mathbf{r}_j^k,t\right)$ is the smoothing parameter and $Z_j$ is a weighting factor to preserve the norm of the state. The spatial distribution of walkers ensures the convergence of the sum in Eq. (14) to the correct probability density function, which is a smooth function in contrast to in Eq. (11). This is important for incorporating the nonlocal correlations which depend on the derivatives of the density. After substituting Eq. (14) into the Hartree equation (10), we obtain:

$$i\hbar \frac{\partial}{\partial t}\varphi_i(\mathbf{r}_i,t) = \left[ -\frac{\hbar^2}{2m}\nabla_i^2 + V_{e-n}(\mathbf{r}_i) + \sum_{j \neq i}^{N} V_{e-e}^{eff}[\mathbf{r}_i - \mathbf{r}_j(t)] + V_{ext}(\mathbf{r}_i,t) \right]\varphi_i(\mathbf{r}_i,t), \quad (15)$$

where:

$$V_{e-e}^{eff}[\mathbf{r}_i - \mathbf{r}_j(t)] = \frac{1}{Z_j} \sum_{k=1}^{M} \int d\mathbf{r}_j V_{e-e}(\mathbf{r}_i - \mathbf{r}_j) \exp\left( -\frac{\left|\mathbf{r}_j - \mathbf{r}_j^k(t)\right|^2}{\sigma_j^k\left(\mathbf{r}_j^k,t\right)^2} \right) \quad (16)$$

is the smeared or effective Coulomb potential. Then, using the TDQMC methodology [19,20], from Eq. (15) we assign separate equation for the guiding wave of each Bohmian walker:

$$i\hbar \frac{\partial}{\partial t}\varphi_i^k(\mathbf{r}_i,t) = \left[ -\frac{\hbar^2}{2m}\nabla_i^2 + V_{e-n}(\mathbf{r}_i) + \sum_{j \neq i}^{N} V_{e-e}^{eff}[\mathbf{r}_i - \mathbf{r}_j^k(t)] + V_{ext}(\mathbf{r}_i,t) \right]\varphi_i^k(\mathbf{r}_i,t), \quad (17)$$

where:

$$V_{e-e}^{eff}[\mathbf{r}_i - \mathbf{r}_j^k(t)] = \frac{1}{Z_j^k} \int d\mathbf{r}_j V_{e-e}(\mathbf{r}_i - \mathbf{r}_j) \exp\left( -\frac{\left|\mathbf{r}_j - \mathbf{r}_j^k(t)\right|^2}{\sigma_j^k\left(\mathbf{r}_j^k,t\right)^2} \right) \quad (18)$$

is the TDQMC effective electron-electron potential, which is easily reduced to a Monte Carlo sum over the walkers:

$$V_{e-e}^{eff}[\mathbf{r}_i - \mathbf{r}_j^k(t)] = \frac{1}{Z_j^k} \sum_{l=1}^{M} V_{e-e}[\mathbf{r}_i - \mathbf{r}_j^l(t)] \exp\left(-\frac{|\mathbf{r}_j^l(t) - \mathbf{r}_j^k(t)|^2}{\sigma_j^k(\mathbf{r}_j^k,t)^2}\right), \qquad (19)$$

where:

$$Z_j^k = \sum_{l=1}^{M} \exp\left(-\frac{|\mathbf{r}_j^l(t) - \mathbf{r}_j^k(t)|^2}{\sigma_j^k(\mathbf{r}_j^k,t)^2}\right) \qquad (20)$$

is the weighting factor. In fact, the effective potential in Eq. (18) describes the weighted nonlocal Coulomb interaction experienced by a given trajectory of the *i*-th electron from the trajectories that belong to the *j*-th electron. Therefore the width of the Gaussian kernel $\sigma_j^k(\mathbf{r}_j^k,t)$ plays the role of correlation length between the electrons. It is seen from Eqs. (19), (20) that when $\sigma_j^k \to 0$ the effective potential tends to the pairwise *e-e* potential of the ultra-correlated case in Eq. (6), while for $\sigma_j^k \to \infty$ the effective potential reduces to the mean-field Hartree potential of Eq. (13). For a given value of $\sigma_j^k$ the wave $\varphi_i^k(\mathbf{r}_i,t)$ which guides the *k*-th walker from the *i*-th electron ensemble experiences the full Coulomb potential due to the *k*-th walker form the *j*-th electron ensemble (for *l=k* in Eqs. (19) and (20)). At the same time $\varphi_i^k(\mathbf{r}_i,t)$ also experiences weighted Coulomb potentials due to the rest of the walkers which represent the *j*-th electron. This is a manifestation of the quantum nonlocality which in our case entangles the trajectories of the Bohmian walkers. The smoothing parameter $\sigma_j^k$ is a natural measure of the electron density of the quantum system and as such it depends on the local density of walkers that can be evaluated by performing adaptive kernel density estimation over the walker distribution. At space locations where the density of walkers is higher the width of the Gaussians in Eq. (14) and Eq. (19) should be smaller in order to compensate for the higher number of Gaussians. Physically, in the regions of higher electron density more intense collisions between the electrons are expected which reduces the correlation length. The mathematical theory of kernel-density estimation suggests a simple formula which relates the width of the Gaussian to the density of particles [28,29]. Accordingly, a pilot density estimate of the walker distribution for the *j*-th electron $\rho_j^k(\mathbf{r},t)$ using kernel density estimation with constant bandwidth σ is calculated first. Then each $\sigma_j^k$ can be estimated through the formula $\sigma_j^k(\mathbf{r},t) = \sigma\sqrt{G_j/\rho_j^k(\mathbf{r},t)}$ where $G_j$ is the geometric mean of the values of $\rho_i^k(\mathbf{r},t)$, for *k*=1,…,M. The bandwidth σ can be determined variationally by minimizing the ground state energy of the system.

It should be noted that effective potentials have already been used to describe quantum systems by classical equations of motion, where the role of the effective potential is to incorporate the quantum effects. Typically these potentials are considered as expansions in $\hbar$ of the quantum partition function. Example of this strategy is the Wigner representation [30] which uses ensembles of classical trajectories in phase space to bring the quantum non-locality effects into the evolution [31,32]. Another example is the Feynman-Hibbs effective potential where Gaussian weighting around the classical trajectory incorporates the quantum effects [33,34]. However, these approximations are essentially quasi-classical.

## 3. Numerical results

To illustrate the contribution of the non-local quantum correlations we calculate within the TDQMC methodology the time-dependent ionization of strongly correlated model system (one-dimensional Helium atom) in an external field. This model atom has proven to be very useful in modeling the interaction of atomic systems with intense ultrashort laser pulses (e.g. in [35]) where modified Coulomb potentials have been employed to avoid numerical complications from the singularity at the origin and between the electrons. Here we assume that the electron-nuclear and the electron-electron interactions are approximated by the following potentials:

$$V_{e-n}(x_i) = -\frac{2e^2}{\sqrt{a+x_i^2}}; \qquad (21)$$

$$V_{e-e}[x_i - x_j^k(t)] = \frac{e^2}{b+\left|x_i - x_j^k(t)\right|}, \qquad (22)$$

where $i=1,2$; $k=1,\ldots,M$, and we have chosen $a=1$ a.u. (atomic units) and $b=1.2$ a.u. in Eqs. (21), (22), respectively. For Helium in a spin-singlet state the two-body wavefunction in Eq. (7) is a symmetrized product of the two one-electron guiding waves, for each couple of walkers which represent different electrons. The time step size in this calculation is 0.1 a.u. while the spatial integration spans 50 a.u. We assign a separate guiding wave $\varphi_i^k(x_i, t=0) = \exp(-x_i^2/\sigma_0^2)$ to each walker ($k$) where $\sigma_o$=0.5 a.u. In order to obtain the walker distribution for the ground state and the corresponding energy we propagate these waves over 400 complex time steps in Eq. (17), together with evolving in real time the Bohmian walkers through Eqs. (7)-(9). The complex time ensures a nonzero velocity of the walkers in Eq. (8) for evolution towards ground state. Using the methodology of Ref. 20, Eq. (41), we found -2.4595 a.u. for the energy of the correlated ground state to be compared with the exact result of -2.4597, while the Hartree-Fock result is -2.4522 a.u. Figure 1 shows the correlation lengths for the two electrons as function of the distance from the core, for 2000 walkers (see the inset). Close to the core where the electron density is higher we have $\sigma_1 \approx \sigma_2 = 0.58 a.u.$ while away from the core $\sigma_1$ and $\sigma_2$ increase by more than an order of magnitude. For time dependent processes the parameters $\sigma_1$ and $\sigma_2$ may need to be updated frequently. However, our calculations reveal that sufficiently accurate results in strong field regime can be obtained using global bandwidth $\sigma_1 = \sigma_2 = 5 a.u.$ instead of performing adaptive kernel density estimation of the streaming data.

Next, we compare the TDQMC result for the time-dependent ionization of the atom with the results from the direct numerical integration of Eq. (1) in 2D configuration space (called here "exact solution"), and from the time dependent Hartree-Fock (TDHF) approximation. For non-relativistic velocities and for wavelengts much larger than the atomic size dipole approximation for the interaction potential in Eq. (17) can be used, $V_{ext}(\mathbf{r}_i,t) = -e\mathbf{r}_i \cdot \mathbf{E}(t)$, where $\mathbf{E}(t) = \mathbf{E}_0(t)\cos(\omega t)$ is the vector of the external electric field. We use linearly polarized electromagnetic pulse with duration eight periods of the carrier frequency and peak intensity in the range 3.5 $10^{14}$ W/cm$^2$ - 1.4 $10^{15}$ W/cm$^2$. Two different carrier frequencies are used with respect to the frequency (0.6117 a.u.) of the first electronic resonance of the model atom. The lower frequency equals 0.136 a.u. while the higher frequency is 1.22 a.u. Figure 2 depicts the time profile of the laser pulse. The direct resonance between the external field and the atom was avoided in these calculations because the resonant excitation of the two-electron system occurs differently if we solve the 2D Schrödinger equation (Eq. (1)) and if the two coupled Schrödinger equations (from Eq. (21)) are integrated. This is because in the former only one of the electrons gets excited in the strongly correlated case, while in the latter both electrons are converted into their excited states. This issue is, however, easily resolved if the two electrons are made distinguishable in the TDQMC method. Since the time dependent ionization of the atom is one of the outcomes most sensitive to the correlation between the electrons, we calculate the ionization by projecting the time dependent state of each electron on its ground state. In Fig.3 the result for the time-dependent ionization obtained from the "effective potential" TDQMC calculation (red line) is presented, compared with the exact result (black line), and with TDHF approximation (blue line). These curves are very close at all times for low frequency and low intensity (Fig. 3a) where the final ionization of the atom is well below 10%. With increasing the laser intensity the encounters between the two electrons become more severe which enhances the ionization in the correlated case in Fig.3b. Figures 3b and 3c show that the predictions of the TDQMC method remain close to the exact results for higher ionizations, while the uncorrelated TDHF result is lower by approximately 20% in Fig. 3c. It is important that in all cases the predictions of the TDQMC method with "effective potential" are much closer to the exact results than the predictions of the ultra-correlated TDQMC method (shown with red dashed lines in Figs. 3-5). For higher laser frequency $\omega=1.22$ a.u., Fig. 4 shows good correspondence between the curves for all intensities. This can be explained by the fact that for high frequencies the electrons are accelerated to much lower velocities than for low frequencies, for the same pulse intensity and width. Therefore the electron-electron collisions are not very strong in this regime, and therefore the results for the ionization in the correlated case are close to the TDHF predictions. In order to confirm these findings we also used different, particle based method to assess the ionization, where it is assumed that a walker is ionized if it travels beyond 10 a.u. from the nucleus. Then, counting the fraction of ionized walkers we plot in Fig.5 the ionization from the TDQMC method compared with the exact result. It is seen that the correspondence between the two curves is fairly good while the ultra-correlated calculation predicts significantly higher ionization in Fig. 5a. It should be noted that the final ionizations in Figs.3, 4 may differ from those in Fig. 5 for the same parameters because the approach used in Figs.3, 4 estimates the ionization in the nearest proximity to the core, unlike that in Fig.5. Having in mind that 1D Helium is one of the most strongly correlated atomic system, our results indicate that the electron-electron dynamic correlation is accurately taken into account within the TDQMC methodology.

## 4. Conclusions

In this paper we have introduced the concept of the "effective potential" into the recently proposed time dependent quantum Monte Carlo approach where quantum dynamics is modeled using ensembles of walkers and time-dependent guiding waves. The effective potential accounts for the nonlocal interaction between the walkers which belong to different electrons and hence correlates different spatial regions of the many-body quantum state. In this way propagation of disturbances in configuration space, which manifests the essential non-locality of the many-body quantum systems, can be described accurately within the TDQMC methodology. The Monte Carlo strategy offers an efficient way to calculate the effective potential, despite its non-local character. Using this approach, the problem for solving the N-body Schrodinger equation is reduced to the numerical solution of number of coupled 3D time-dependent Schrodinger equations for the individual guiding waves, and separate equations for the motion of the corresponding walkers. These equations do not involve calculation of multiple integrals over wavefunctions which provides a very good scaling as compared to other quantum many-body methods. Our calculations show that the ionization of 1D Helium atom in strong laser field is correctly described by the "effective potential" TDQMC method. For low carrier frequency the ionization of the correlated electrons is enhanced in TDQMC as it is in the exact solution, while the time-dependent Hartree-Fock method does not predict such enhancement. For higher frequencies the correlated and uncorrelated results are very close. The results of this work can be readily extended to the case where both the electron and the nuclear motion are treated quantum-mechanically. Since the nuclear mass is much higher the parameter $\sigma$ for the nuclei is expected to be much smaller than that for the electrons.

**Acknowledgments**

The author gratefully acknowledges support from the National Science Fund of Bulgaria under contract WUF-02-05.

# References

[1] M. Hentschel, R. Kienberger, Ch. Spielmann, G. A. Reider, N. Milosevic, T. Brabec, P. Corkum, U. Heinzmann, M. Drescher and F. Krausz, Nature **414**, 511 (2001).

[2] G. Sansone, E. Benedetti, F. Calegari, C. Vozzi, L. Avaldi, R. Flammini, L. Poletto, P. Villoresi, C. Altucci, R. Velotta, S. Stagira, S. De Silvestri, M. Nisoli, Science **314**, 443 (2006).

[3] M. Drescher, M. Hentschel, R. Kienberger, M. Uiberacker, V. Yakovlev, A. Scrinzi, Th. Westerwalbesloh, U. Kleineberg, U. Heinzmann, and F. Krausz, Nature **419**, 803 (2002).

[4] A. L. Cavalieri, N. Muller, Th. Uphues, V. S. Yakovlev, A. Baltuska, B. Horvath, B. Schmidt, L. Blumel, R. Holzwarth, S. Hendel, M. Drescher, U. Kleineberg, P. M. Echenique, R. Kienberger, F. Krausz, and U. Heinzmann, Nature **449**, 1029 (2007).

[5] E. Runge and E. K. U. Gross, Phys. Rev. Lett. **52**, 997 (1984).

[6] M. A. L. Marques and E. K. U. Gross, Annu. Rev. Phys. Chem. **55**, 427 (2004).

[7] F. Furche and K. Burke, in Annual Reports in Computational Chemistry, edited by D. Spellmeyer (Elsevier, Amsterdam, 2005), vol. 1

[8] M. Petersilka, U. J. Gossmann, and E. K. U. Gross, Phys. Rev. Lett. **76**, 1212 (1996).

[9] N. T. Maitra, K. Burke, and C. Woodward, Phys. Rev. Lett. **89**, 023002 (2002).

[10] A. Dreuw, J. Weisman, and M. Head-Gordon, J. Chem. Phys. **119**, 2943 (2003).

[11] N. T. Maitra, J. Chem. Phys. **122**, 234104 (2005).

[12] M. van Faassen and K. Burke, Chem. Phys. Lett. **431**, 410 (2006).

[13] J. Zanghellini, M. Kitzler, C. Fabian, T. Brabec, and A. Scrinzi, Laser Phys. **13**, 1064 (2003).

[14] T. Kato and H. Kono, Chem. Phys. Lett. **392**, 533 (2004).

[15] M. Nest, R. Padmanaban, and P. Saalfrank, J. Chem. Phys. **126**, 214106 (2007).

[16] T. Klamroth, J. Chem. Phys. **124**, 144310 (2006).

[17] N. Rohringer, A. Gordon, and R. Santra, Phys. Rev. A **74**, 043420 (2006).

[18] M. Nest, J. Theor. Comput Chem. **6**, 563 (2007).

[19] I. P. Christov, Opt. Express **14**, 6906 (2006).

[20] I. P. Christov, J. Chem. Phys. **127,** 134110 (2007).

[21] B Hammond, W. Lester, and P. Reynolds, *Monte Carlo Methods in Ab Initio Quantum Chemistry* (World Scientific, Singapore, 1994).

[22] W. M. C. Foulkes, L. Mitas, R. J. Needs, and G. Rajagopal, Rev. Mod. Phys. **73**, 33 (2001).

[23] R. E. Wyatt, *Quantum Dynamics with Trajectories* (Springer, Berlin, 2005).

[24] P. R. Holland, *The Quantum Theory of Motion* (Cambridge University Press, Cambridge, 1993).

[25] M. Namiki, *Stochastic Quantization* (Springer, Berlin, 1992).

[26] D. Bohm and J. P. Vigier, Phys. Rev. **26**, 208 (1954).

[27] E. Nelson, Phys. Rev. **150**, 1079 (1966).

[28] I. S. Abramson, Ann. Statist. **10**, 1217 (1982).

[29] B. W. Silverman, *Density Estimation for Statistics and Data Analysis,* Monographs on Statistics and Applied Probability (Chapman and Hall, London, 1986).

[30] E. Wigner, Phys. Rev. **40**, 749 (1932).

[31] W.H. Miller, PNAS **102**, 6660, (2005).

[32] A. Cuccoli, A. Macchi, V. Tognetti, M. Neumann, R. Vaia, Phys. Rev. B **45**, 2088 (1992).

[33] R. P. Feynman and A. R. Hibbs, *Quantum Mechanics and Path Integrals* (McGraw-Hill, New York, 1965).

[34] R. P. Feynman and H. Kleinert, Phys. Rev. A **34,** 5080 (1986).

[35] R. Grobe and J. H. Eberly, Phys. Rev. Lett. **68**, 2905 (1992).

**Figure captions:**

**Figure 1**. (Color online) Dependence of the correlation lengths of the two electrons on the distance from the core. The inset shows the walker distribution for the ground state.

**Figure 2**. Time dependence of the electric field used in the calculations, for carrier frequency $\omega=0.136$ a.u.

**Figure 3.** (Color online) Time dependent ionization for 1D model Helium atom in external field with low carrier frequency $\omega=0.136$ a.u. and amplitude: $E_o=0.10$ a.u.- plot (a), $E_o=0.15$ a.u.- plot (b), and $E_o=0.2$ a.u.- plot (c). The different colors correspond to different approximations: black line-exact result, red line-TDQMC result, blue line-TDHF result, red dashed line-ultra correlated TDQMC. Projection on the ground state is used to calculate the ionization.

**Figure 4**. (Color online) Time dependent ionization for 1D model Helium atom in external field with high carrier frequency $\omega=1.22$ a.u. and amplitude: $E_o=0.10$ a.u.- plot (a), $E_o=0.20$ a.u.- plot (b), and $E_o=0.30$ a.u.- plot (c). The different colors correspond to different approximations: black line-exact result, red line-TDQMC result, blue line-TDHF result, red dashed line-ultra correlated TDQMC. Projection on the ground state is used to calculate the ionization.

**Figure 5**. (Color online) Time dependent ionization for 1D model Helium atom – particle-based approach. The walker is assumed to be ionized if it travels beyond 10 a.u. from the nucleus. Black line -exact result, red line -TDQMC result, red dashed line-ultra correlated TDQMC. Plot (a)- low frequency $\omega=0.136$ a.u., $E_o=0.20$ a.u., plot (b) – high frequency $\omega=1.22$ a.u., $E_o=0.30$ a.u.

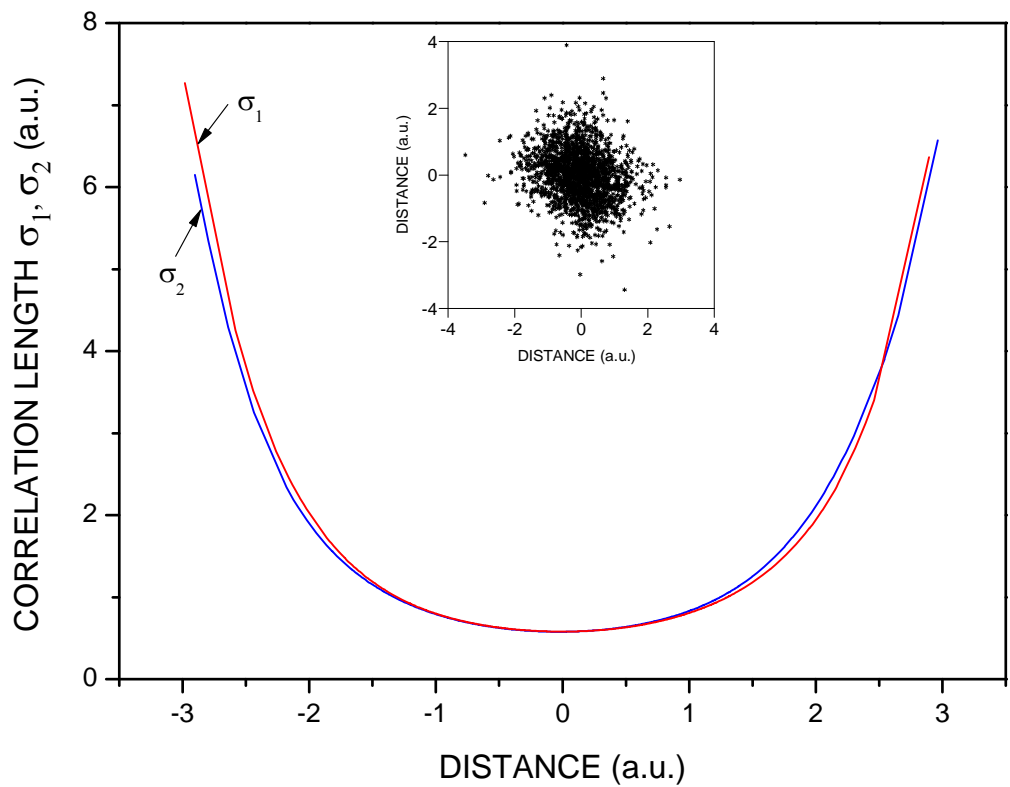

I. Christov, Fig.1

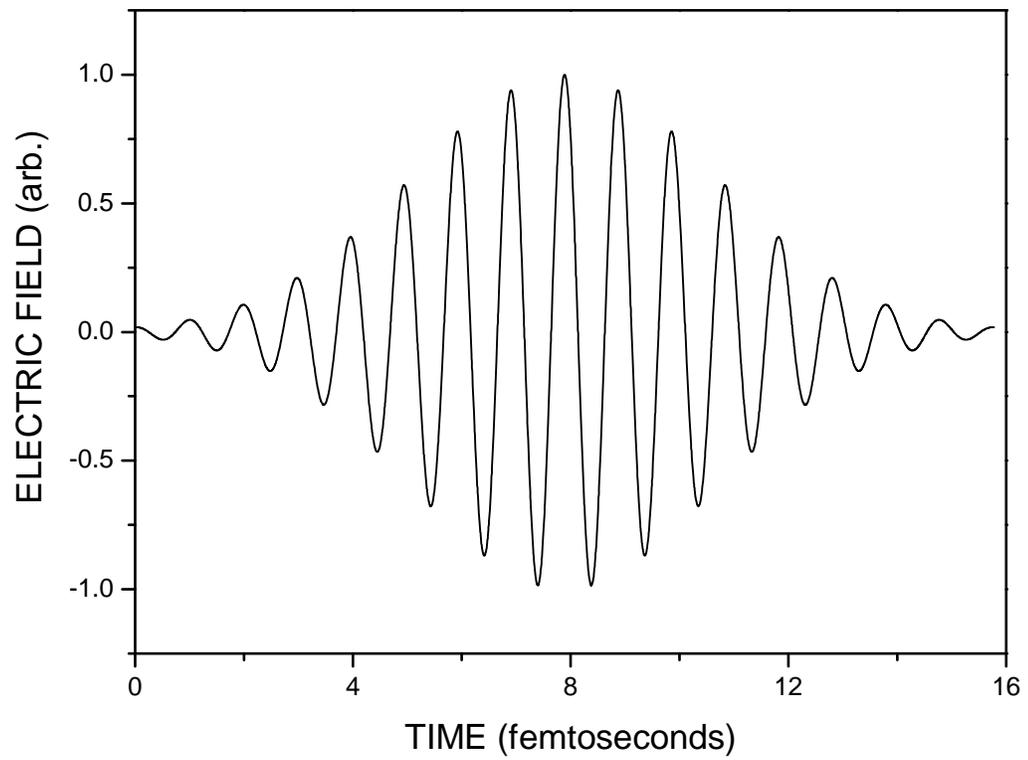

I. Christov, Fig.2

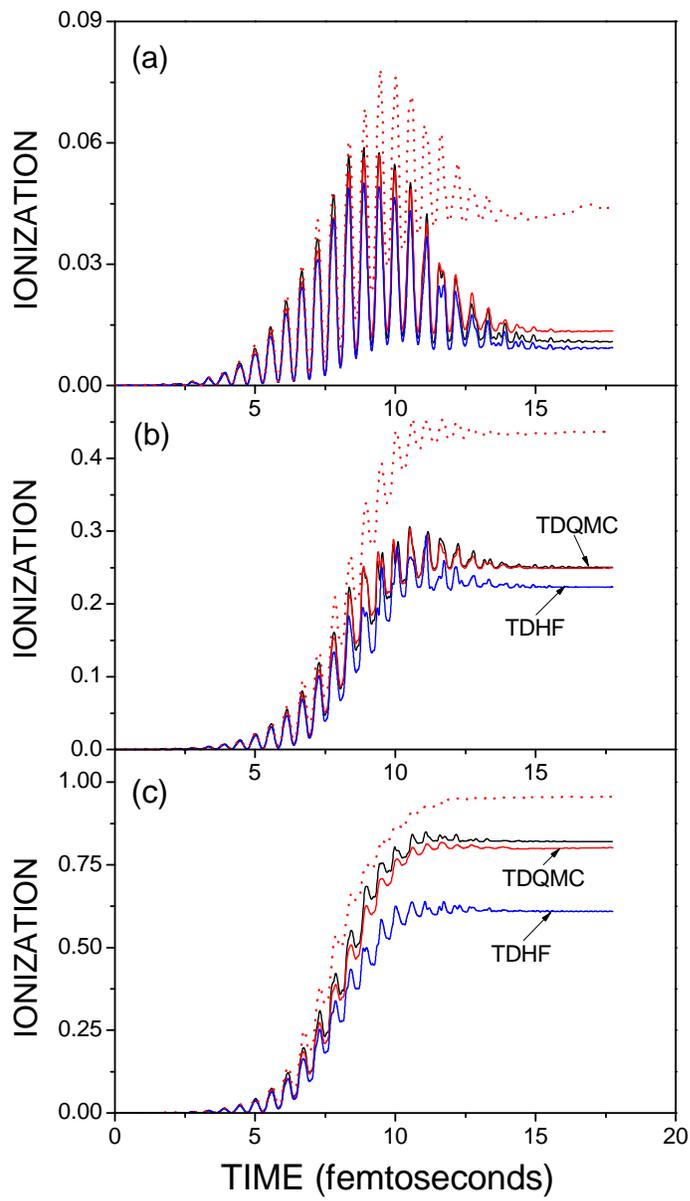

I. Christov, Fig.3

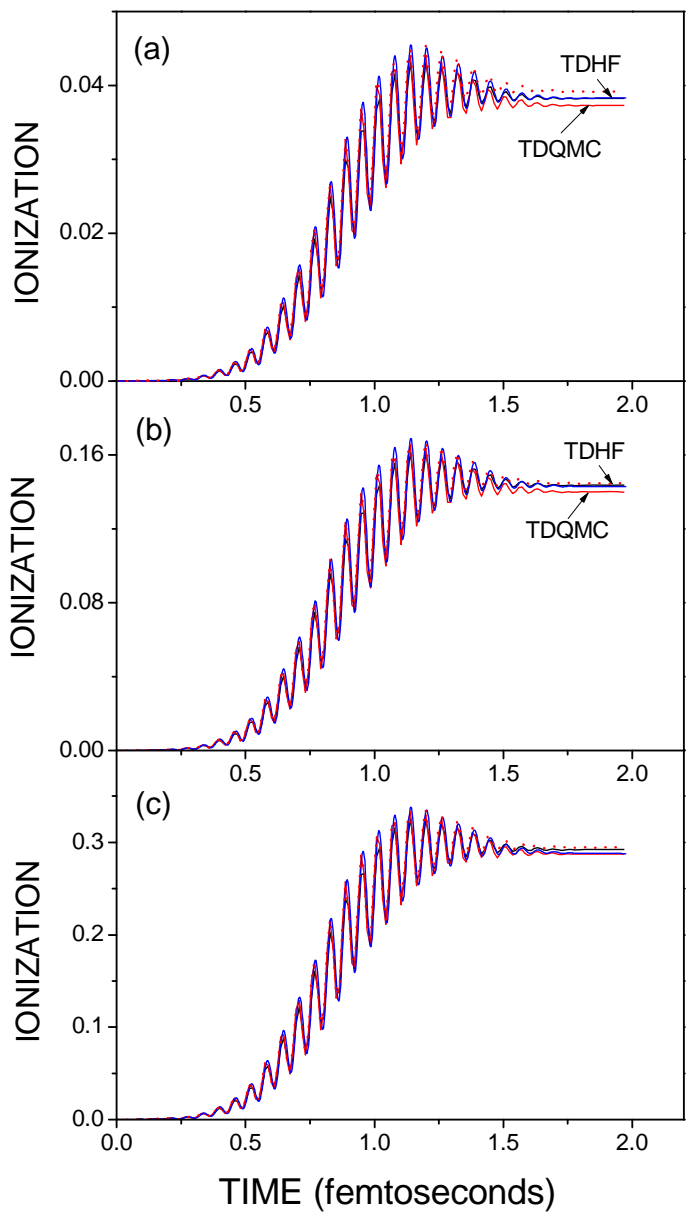

I. Christov, Fig.4

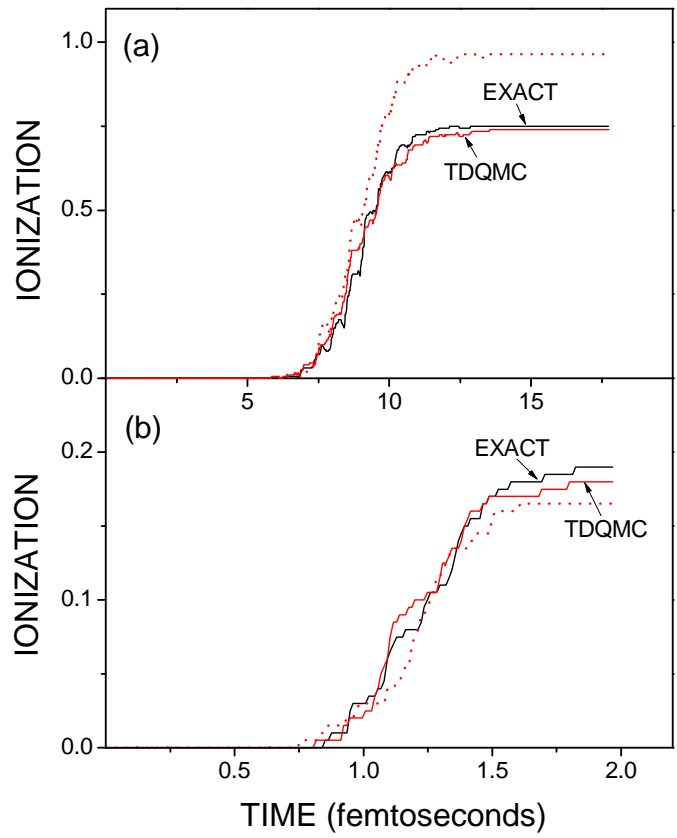

I. Christov, Fig.5